# Comprehensive review for common types of errors using spreadsheets

Ali Aburas
School of EECS
Oregon State University
aburasa@eecs.oregonstate.edu

**Abstract**

Thanks to their flexibility and capability to perform different tasks and organize data in the best form and format, spreadsheets are widely used in different organizations and by different end users. Many business organizations rely on spreadsheets to fulfill their various tasks. On the other hand, the number of spreadsheets that contain errors are very high, thus researchers have developed different tools aimed at the prevention, detection, and correction of errors in spreadsheets. This research work is a comprehensive review that describes and classifies approaches on finding and fixing errors in spreadsheets. The paper discusses up-to-date research work approaches in terms of definition, how they work, and kinds of errors they can find in spreadsheets. The paper looks also for the kinds of errors that end users commonly make in spreadsheets.

## 1 Introduction

By 2012, there will be around 90 million American workers using computers. In addition, there will be over 55 million workers who will do programming based on spreadsheets, followed by databases with different levels of complexity to help solve difficulties in their jobs. On the other hand, there will be fewer than 3 million professional programmers in the United States itself [8].

Spreadsheet programs are probably the most well-known programming systems among end-user programmers. The reason that they are prevalent is that end users can easily use them without requiring any type of training in programming [1]. Most end-user programmers use spreadsheets to program [1]. These programmers might be teachers, secretaries, or even children. In general, they often write programs to support their work or for personal use, and they do not have a solid background in computer science [1]. For example, a teacher can create a spreadsheet file that tracks all students' scores during a school year.

On the one hand, the flexibility afforded by spreadsheet programs allows end users and many organizations to use spreadsheets as decision support tools. On the other hand, this flexibility makes it easy to create errors in spreadsheets or introduce new errors in existing ones. Each year millions of end users, such as managers and professionals, create hundreds of millions of spreadsheets to help them in making critical decisions [2].

Most end-user programmers are self-taught and pay inadequate attention to the dangers of errors in their spreadsheet files [4]. A study reports that 90% or more of real-world spreadsheets contain errors [5]. The implications of a small number of errors in a spreadsheet are very serious, and the consequences could be very costly. For example, a Florida-based construction company lost almost a quarter of a million dollars by relying on an incorrect spreadsheet [3].

Varieties of approaches have been applied to prevent, detect, and remove errors from spreadsheets. This research paper is a comprehensive review of several different approaches to detect and fix errors in spreadsheets. Section 2 looks at different types of errors that end users make in their spreadsheets. Section 3 presents different approaches for preventing errors in spreadsheets by generating spreadsheets from user-defined templates and gives a variety of effective approaches for detecting and correcting spreadsheet errors. Then, Section 4 contains a full discussion of all approaches and potential future works. Finally, a conclusion statement about the reviewed approaches is given.

**2 Types of errors in spreadsheets**

The distinction between different classes of spreadsheet errors is important since they allow researchers to understand cause, frequency, and prevention of the spreadsheet errors. However, there is no accepted classification of spreadsheet errors because different classifications vary by purpose, and a classification that is suitable for one purpose may not be suitable for another [4]. For example, in a lab experiment, entering an incorrect formula with a typing error might be possible to observe. However, when discovered during a field audit, that same error might be interpreted as an error in logic [44].

Researchers have classified spreadsheet errors into two main types: *quantitative* and *qualitative* [2]. Quantitative errors are wrong results in the spreadsheet. In contrast, qualitative errors are faults that do not produce an immediate error, but they degrade the quality of the spreadsheet and might lead to quantitative errors during later updates. In addition, quantitative errors have been further divided into *mechanical*, *logic*, and *omission errors*. The three types were defined as follows:
1. *Mechanical errors* are simple mistakes that end users make due to negligence or distractions. For example, mistyping a number or a reference, or referring to a wrong cell address [1].
2. *Omission errors* occur when something is accidentally left out of the spreadsheet model or the situation to be modeled is misunderstood. For example, references to corresponding input data in the output section are omitted from the model [5]. These types of errors are very dangerous because they have low detection rates [2].
3. *Logic errors* arise when the end user applies the wrong formulas or algorithms for solving a problem. These types of errors usually need domain-specific knowledge to find and correct [1].

Rajalingham, Chadwick, and Knight developed a more elaborate classifications approach [6]. The approach used a binary tree, and for the first time, non-human generated errors were considered. However, a revision was made to these classifications, and some repetitious groups were removed by avoiding the distinction between end-user and developer-created errors [7]. The revised spreadsheet errors are classified into *application-identified errors* and *developer/user-identified errors*. However, I am going to consider only the *developer/user-identified errors* as within the scope of this paper.
1- *Qualitative errors*
   A. *Structural errors* occur as a result of flaws in the design or layout of the model that cause confusion. They are further divided into *hidden* and *visible errors*. The *structural visible errors* represent errors that do not require an examination of formula, such as incorrect or ambiguous headings. However, the *structural hidden errors* represent errors that require an examination of formula. For example, if an end user formats cells to one digit to the right of the decimal, the spreadsheet will round any values that might have greater precision. For example, the sum of 1.44 and 1.44 will round to 2.9 from 2.88. Such additions will appear incorrect and could cause confusion [5].
   B. *Temporal errors* occur when a spreadsheet uses data that has not been updated. As a result, these types of errors can lead to unreliable decisions or interpretations of the real situation [5]. For example, if there is a spreadsheet that contains a formula that uses an exchange rate for two different currencies, the calculation of the formula produces a value that is invalid if the exchange rate has become invalid due to time lapse [5].
2- *Quantity errors*
   A. *Reasoning errors* involve creating the incorrect formula due to choosing the wrong algorithm or writing the wrong formula to implement the algorithm. Reasoning errors can be divided into *domain knowledge* and *implementation errors*.

i. *Domain knowledge errors* occur when the end user does not have the skills to identify and analyze business functions that are appropriate for modeling with a spreadsheet. Domain knowledge errors can be divided into *real-world knowledge* and *mathematical representation errors*.
      a. *Real-world knowledge errors* occur when the end user creates a wrong formula by selecting the wrong algorithm. For instance, dividing a leap year by 365 instead of 366 [5].
      b. *Mathematical representation errors* occur when the end user constructs an incorrect or inaccurate formula to implement a correct algorithm. For instance, an error occurs when the formula that calculates percentage is incorrectly written because of missing brackets [5].
   ii. *Implementation errors* occur due to a lack of knowledge of the functions and capabilities of the spreadsheet system in use. Implementation errors can be divided into *syntax errors* and *logic errors*.
      a. *Syntax errors* occur when the spreadsheet software cannot perform a formula that contains strange characters or symbols. However, the spreadsheet systems easily detect these types of errors and indicate that an error has occurred [5].
      b. *Logic errors* occur due to the lack of understanding of the features of the spreadsheet software in use. For example, most end users think that the spreadsheet software, such as Excel, will automatically alter row and column references wherever they copy a formula, but sometimes the spreadsheet cannot alter the formula's references because it contains absolute references. This is called the relative and absolute copy problem [5].
   iii. *Accidental errors* are mistakes caused by carelessness, such as typos. These errors can be spotted and corrected immediately by the end user. Otherwise, they could lead to incorrect values in other cells. For example, new rows are added to spreadsheets, but the formula that calculates the total of the rows has not been updated to include these new rows [5].

However, this classification still has limitations as its categories occasionally overlap. For example, an error caused by a lack of domain knowledge could accidentally occur during insertion [4]. Another limitation is that it is sometimes hard to recognize the cause of an error enough to accurately classify it. For example, a formula giving incorrect results could be due to a lack of domain knowledge, or due to a typing mistake [4].

The high error rates in spreadsheets have encouraged many researchers to conduct different research works to support end users in creating reliable spreadsheets and improve the quality of spreadsheets with less effort. In the following section, I am going to review different approaches aimed at helping end users in designing, testing, and debugging their spreadsheets.

**3 Types of approaches**

Spreadsheet programs lack any higher-level abstractions, which makes it very difficult to maintain spreadsheets that implement complex models [1]. As a result, end users easily make errors when they attempt to create or reuse complex spreadsheets. Researchers have developed different approaches that engage different software engineering activities to improve the quality of spreadsheets. These approaches are generally focused on the prevention, detection, and removal errors from spreadsheets [6]. In this section, I will illustrate some error prevention, auditing, testing, and automatic consistency checking approaches to detect and remove errors.

**3.1 Preventing errors**

Spreadsheets are error-prone because they do not offer any type of abstractions that allows separating data from computation, which is the underlying spreadsheet model. In other words, spreadsheet systems allow updating the data and the model on the same spreadsheet, and they do not impose any types of restrictions on the kinds of updates that can be carried out [1]. This problem has led to the development of an approach that allows end users to create correct structures of spreadsheets, which are called templates, and then generate spreadsheets that conform to the template.

As a result, researchers have developed a specification language called Visual Template Specification Language (ViTSL) [10]. ViTSL is a visual language for defining the underlying model of a spreadsheet in the form of a template and the ways it evolves. Then, the template can be imported to another related system called Gencel, which generates spreadsheets based on the template with customized update operations [11]. By applying these two approaches, all underlying model updates are performed at the template level, and only data changes are allowed within the Gencel system.

There are two significant constructs for creating a template in ViTSL. A vex group indicates that a group of consecutive rows can be repeated in the vertical direction. Likewise, a hex group indicates that a group of consecutive columns can be repeated horizontally. For example, consider the spreadsheet in Figure 1(a). It shows the budget template in the ViTSL system, where a template is developed, which defines headers, data cells, and computations (formulas). The horizontal dots after column D indicate columns B, C, and D can be horizontally expanded, and the vertical dots indicate that row 3 can be repeated. Figure 2(b) shows an instance of the budget template in the Gencel system; the end user can perform insertion and deletion of rows and columns through the menu bar on the right.

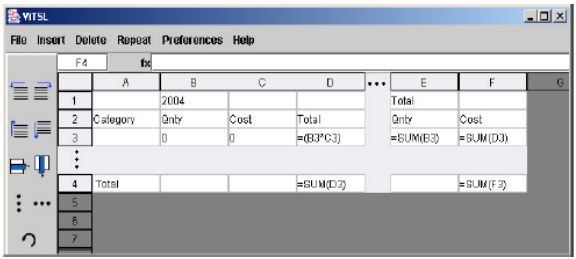
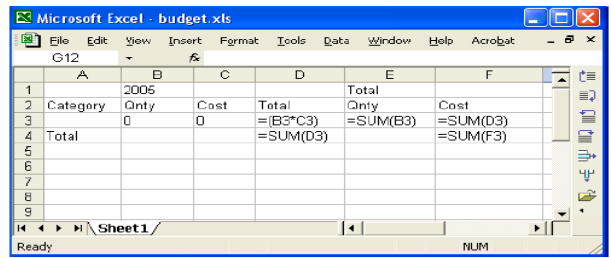

(a) Budget template created in the ViTSL editor     (b) Gencel budget spreadsheet

Figure 1: Budget template created with the ViTSL/Gencel. [10]

Since Gencel automatically performs all the necessary formula generation and spreadsheet structure modifications during the evolution of a spreadsheet, the end user can only concentrate on entering the data within the spreadsheet and does not need to worry about updating formulas.

Implementing this approach as an extension of Excel prevents different kinds of errors, such as omission errors (omitted cells in aggregations), reference errors (wrong references), type errors (operands have illegal types) [11], and structural errors that can occur due to a flaw in the design of a spreadsheet model.

Domain experts or professional programmers could create different templates, and these templates can be distributed to others for use. However, the creation of templates from scratch limits the applicability of the approach. Thus, Parcel has been developed for extracting templates automatically [12]. This tool infers the underlying model from existing spreadsheets, which allows different organizations to import legacy spreadsheets into the ViTSL and Gencel framework.

The ViTSL/Gencel approach is limited in its scope since it relies on vertical and horizontal repetition of related rows and columns to express structures of spreadsheets. As a result, an object-oriented extended template approach has been developed that allows for the capture of the underlying business structure of a spreadsheet using object-oriented classes [42]. This is called a ClassSheet template because it defines classes together with their attributes and aggregation relationships.

Consider, for example, that the so-called income sheet in Figure 2-left is shown as a ClassSheet and it consists of a list of numeric data values, which are summed up in a separated cell. From an object-oriented point of view, we can observe that it consists of a class pattern of two classes, Income and Item, representing a one-dimensional aggregation structure. Hence, ClassSheets consist of a list of attribute definitions grouped by classes and are arranged on a two-dimensional grid. The class names are set in boldface, such as Income and Item, in contrast to attribute names and labels, which are set in normal face, such as total and value. In addition, colored borders are used to represent the different classes within a ClassSheet. References to other cells are defined by using attribute names, as shown in the SUM formula in the example. This type of notation helps end users prevent the design of incorrect data computations. In summary, ClassSheets specification can be translated into an equivalent ViTSL template in which spreadsheet applications can be generated. Similarly, a UML representation may be derived from a ClassSheet (see Figure 2-right).

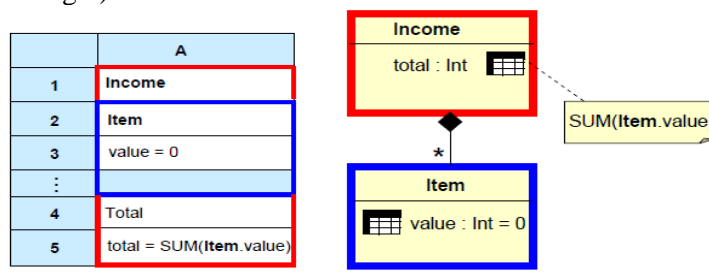

Figure 2: A simple one-dimensional ClassSheet. [42]

Designing such models is time-consuming and requires expertise. Therefore, a technique that can automatically infer ClassSheets from spreadsheets using object-oriented classes and database concepts has been provided [43].

### 3.2 Detecting errors

Prevention approaches require additional programming in a new language that end users must learn, which can be very time consuming. Therefore, much research has focused on the detection and removal of errors. The detection of spreadsheet errors has been mainly focused on three different approaches: auditing, testing, and automatic checking [13].

### 3.2.1 Auditing

Spreadsheets are easy to build, but their correctness is hard to check. It is almost impossible to check the correctness of a spreadsheet without understanding the purpose of the spreadsheet, which requires understanding a formal definition of its computations [14]. In addition, when an existing spreadsheet needs to be modified, its underlying model must be understood before any changes can be made. The main obstacle that end users face in understanding spreadsheets is the difficulty in knowing how each cell depends on other cells. However, end users must examine each cell that contains a formula one by one to understand the model of a spreadsheet. As a result, this process is very tedious and error-prone.

Researchers have developed different approaches to make the connections between the cells visible through visual depiction. They have developed auditing and visualization approaches to make spreadsheets easier to comprehend and errors easier to detect. These approaches deal with analyses of spreadsheets on the formula level by recognizing and coloring similar groups of cells based on formula similarities they contain [15]. In the following, some auditing approaches that visualize the deep structure of spreadsheets for the purposes of error detection and comprehension of spreadsheets will be described.

Different techniques have been presented that make the hidden dataflow structure visible and accessible [16]. These techniques allow end users to visually interact with the hidden structures and to understand them better. In addition, they visualize spreadsheet structures at varying levels of connections. Essentially, the first level visualizes both incoming and outgoing cells of a cell, which is called *transient local view*. Then, the entire structure of a spreadsheet related to one cell is visualized, which is called *static global view*. These visualizations can be animated, which automatically generates an animated presentation to reveal the structure of the spreadsheet. The animated global explanation starts at initial cells that contains data; then joins and split at intermediate cells; and lastly ends up at the terminal cells, which contain the result of computation.

However, the drawback of these techniques is that spreadsheets' structures cannot be visualized area by area, resulting in a cluttered appearance and them being difficult to follow. Therefore, more new spreadsheet visualization approaches that decompose large spreadsheets into two levels of abstractions, *logical regions* and *semantic classes*, have been developed.

Other researchers consider spreadsheets the result of a copy, paste, and modify process [15, 17]. As a result, they define methods for classifying spreadsheets' areas into logical regions and semantic classes. Logical regions construct an abstract presentation of a spreadsheet by assigning its cells based on the similarity between the formulas in one of the following equivalence classes:
1- *Copy equivalence* exists when the formulas are identical.
2- *Logical equivalence* exists when the formulas differ only in constant values and absolute references.
3- *Structural equivalence* exists when the formulas contain the same operation in the same order.

All relative references are compared in the R1C1-notation. A simple spreadsheet illustrates equivalence classes in Figure 3. We can see B1 and B2, and D1 and D2 are surrounded by thick borders, indicating two different logical regions based on copy equivalence. Meanwhile, the grey shaded area indicates a logical area of logical equivalence. Lastly, cells B1, B2, C1, C2, D1, and D2 are structurally equivalent [18].

Figure 3: Example spreadsheet, formula view. [18]

However, if a very large spreadsheet is analyzed, there will be many logical areas, and they cannot be displayed on a single screen. Therefore, there is a need for a higher level of abstraction that groups the cells into larger areas. Semantic classes are based on logical areas but offer a higher level of abstraction and contain blocks of cells that are repeatedly used throughout the spreadsheet, i.e. similar cells with similar neighbors. As a result, semantic classes can deal with regular large spreadsheets.

Other researchers implement different techniques for combining replicated formulas, which most spreadsheets contain, into regions and use them for a testing methodology known as "What You See Is What You Test" (WYSIWYT), *Sub-section 3.2.2*, to decrease testing and computational effort and improve the efficiency of dataflow testing approaches [19]. However, these regions are not shown to the end users testing.

These techniques infer regions in two steps. The first step defines two cells as formula similar if their formulas are formula equivalent and any one does not contain any references to other cells. The second step involves grouping similar cells into regions. Region-inference techniques take into account that regions are not necessarily rectangular. Therefore, large regions can be found to improve error detection and decrease computational effort.

Three different techniques for inferring regions yield three spatial relationships regions:
1- *Discontiguous regions* can be identified by merging all the cells in a spreadsheet that are formula similar, and these regions yield the most general concept of what constitutes a region. In addition, regions can be discontiguous, containing cells that are not neighbors.
2- *Contiguous regions* can be identified by comparing all the cells with their neighboring cells and merging them into regions if their formulas are formula similar.
3- *Rectangular regions* can be found by merging all the cells into regions with their above or below cells if their formulas are formula similar, and then comparing all the identified regions to the regions on either side of them, and merging the adjacent regions with formula similar formulas with the same height.

A controlled experiment investigated the feasibility of this approach and it showed that the discontiguous regions technique (D-Regions) found fewer regions than the other techniques. Thereby, D-Regions require less testing effort and fewer interactions in the spreadsheet [19].

Visualization approaches can be very helpful for end users in situations that required comprehending and interacting with data. They also can be very helpful in detecting omission errors, such as a missing reference or a wrong operation. These errors are not obvious on the value level of the spreadsheet. However, a major limitation of these approaches is that they do not give any feedback to the end user on the spreadsheet correctness.

### 3.2.2 Checking, testing, and debugging

Testing can expose faults within programs. That is why professional programmers test their programs to gain confidence that it works as expected, and they spend a large amount of time identifying and correcting errors within their programs [9]. Spreadsheet systems do not support any type of systematic testing of spreadsheets. As a result, end users try to do simple tests by changing formulas and looking at the immediate results. However, these tests allow them to build a high level of confidence in the correctness of their spreadsheets. Studies show that end users are often overconfident in the correctness of their spreadsheets, and they assume that their spreadsheets are correct [20]. Therefore, research on testing and verification aims for the spreadsheet systems to help end users manage their overconfidence.

WYSIWYT is a methodology for testing spreadsheets. It allows end users to systematically test individual cells in their spreadsheets, and it keeps track of which cells in the spreadsheet have been tested. Whenever the end user enters values into the spreadsheet, the end user marks the output cells, which contain formulas, with a √ if the output is correct and an *x* if incorrect. In addition, WYSIWYT uses definition-use (du) adequacy, which executes all the data dependencies between cell formulas caused by references to other cells for measuring the level of the testedness of the spreadsheet [21]. The testedness is shown to the end user by coloring the borders of the noninput cells from untested (red color) to tested (blue color) and through a progress bar, which ranges from 0 to 100 percent.

The benefits of the WYSIWYT have been shown empirically [23], and the results show that the WYIWYT improves end users' effectiveness, which was measured by the percentage of du-associations covered by the test cases and the number of faults detected, and efficiency, which was measured by redundancy and speed, in testing even without training in testing theory. A study compared the Ad Hoc and WYIWYT subjects' performances on two different types of problems; Table 1 shows the results. It is clear that WYSIWYT subjects' testing performance, in terms of efficiency and effectiveness, in the first problem was greater than the Ad Hoc subjects'. In addition, the WYSIWYT subjects improved their testing on the second problem. It also shows that Ad Hoc subjects ran redundant test cases, which means that they executed multiple test cases that executed the same du-association and did not increase testing effectiveness more than WYSIWYT subjects. In conclusion, these results show that subjects using the WYSIWYT performed more effective testing than the Ad Hoc subjects, and they were more efficient testers than the Ad Hoc subjects.

|  | Problem 1 | | | Problem 2 | | |
| --- | --- | --- | --- | --- | --- | --- |
|  | Tested | # Tests | Redundant | Tested | # Tests | Redundant |
| Ad Hoc | 69.0% | 13 | 51.3% | 71.6% | 22 | 56.3% |
| WYSIWYT | 82.7% | 20 | 11.1% | 97.8% | 18 | 7.7% |

Table 1: Learning Effects: compare the medians of Problem 1 and Problem 2. [23]

By focusing on dependencies between cells, WYSIWYT helps end users detect a wide range of faults related to formulas, including reference, operator, and logic errors [23]. However, testing a spreadsheet generally requires creating many appropriate test cases to guarantee detecting all faults, which the end user may be unable to generate manually. Therefore, two automatic test-case-generation mechanisms have been developed for spreadsheets: "Help Me Test" (HMT) [24], which has been integrated into WYSIWYT, and AutoTest [25].

HMT basically implements two different algorithms to generate test cases. *Randomly chosen inputs* randomly assigns values to the input cells then checks whether any du-association is exercised or not. *Chaining approach* finds a set of values of input cells that cause one or more links in du-association to be executed. The basic idea of the chaining approach is to identify a sequence of cells and branches to be executed prior to execution of the formula that is being tested. An empirical study's results show that HMT is highly effective and efficiently generates test cases [24].

AutoTest allows the end user to select a cell that contains a formula to be tested. The system generates a set of test values for all the input cells that the formula depends on and presents the values together with the corresponding output of the selected cell [25]. The end user marks the generated test case as valid if its output matches the expected output of the formula or flags it to indicate that the computed output value is incorrect and the formula is faulty. If the formula being tested is faulty, it is shaded red and can later be modified. In addition, if the end user cannot decide whether the output is correct or not, he or she can ignore the generated test case. As testing continues, a progress bar shows the degree of testedness of the spreadsheet.

AutoTest generates constraints and automatically generates test cases by backward propagation and solution of constraints on cell values. These constraints are obtained from the formula of the cell that is being tested. A comparative evaluation of AutoTest against the "Help Me Test" shows that AutoTest is faster and produces test suites that give better DU coverage [25].

Although verification and testing indicate the existence of errors, debugging is the process of locating and correcting errors. End users face challenges to debug a spreadsheet. Most of the time, end users cannot figure out the cause of a particular failure in their spreadsheets. On top of that, if end users are able to identify the cause of a failure, it is not always easy to make changes because of the complexity of the formulas, and there is the danger of introducing more errors while trying to correct an existing one. These issues led to the development of a semi-automatic debugger for spreadsheets called GoalDebug [27].

GoalDebug allows the end users to indicate an expected value or range of values for cells that give incorrect results and get a list of change suggestions to the spreadsheet's formulas to achieve the expected value in the target cells. The system uses dynamic slicing techniques to generate the change suggestions and propagates expected values backward across formulas. After that, the system uses heuristic techniques to rank the changes from most likely to least likely. The end user can select, explore, or ignore the change suggestions, and if any one is selected, the selected one can be applied automatically to the formulas; thus, there is no need for a formula to be manually edited.

GoalDebug helps end users lower the cost of finding and correcting different kinds of errors such as omission, type, and reference errors. Its usefulness is limited if the end user provides incorrect values. The effectiveness of automatic test-case generators has been evaluated by measuring the quality of test suites using mutation testing.

Mutation testing involves deliberately inserting faults into the original program and creates a set of faulty programs called mutants. The goal in mutation testing is to measure the effectiveness of a test suite in terms of its ability to detect all the seeded faults, which is called mutation adequate. To measure the quality of a given test suite, the mutants are executed against the input test suite to see if the inserted fault can be detected. If the test suite detects the mutant, it is assumed to be effective at detecting faults in the original program. Abraham and Erwig develop a suite of mutation operators for spreadsheets, which can be used to seed spreadsheets with errors [26] and demonstrate their use in the evaluation of AutoTest [25] and GoalDebug [27]. The mutation operators were chosen to replicate the errors commonly made by end users and to help the tester develop effective tests or locate weaknesses in the test data used for the spreadsheet. The result of using mutation-testing approach shows that GoalDebug ranks the correct changes among the top five in 80% of the test cases with adapting new ranking heuristics compared to 67% of the cases with the old version of the system. In addition, the result shows that a modified AutoTest approach that generates test cases for all solvable constraint sets is successful in detecting almost all the mutants. As a result, evaluation shows some faults were discovered by mutation-adequate test suits that were not discovered by generated test suites that satisfy the du-adequacy criterion.

Another approach implemented to detect faults in spreadsheets' formulas is called assertions [28]. In this approach, assertions are attached to spreadsheet cells to identify a range of numerical values that can be entered in those cells. In other words, assertions protect cells from bad values and work as guards as they guard the correctness of the spreadsheet cells [29]. This approach, shown in Figure 4, allows end users to explicitly enter a range of numerical values that a spreadsheet cell can have, which are called *user assertions (USA)*. The system then propagates user assertions through formulas to product *system-generated assertions (SGA)*.

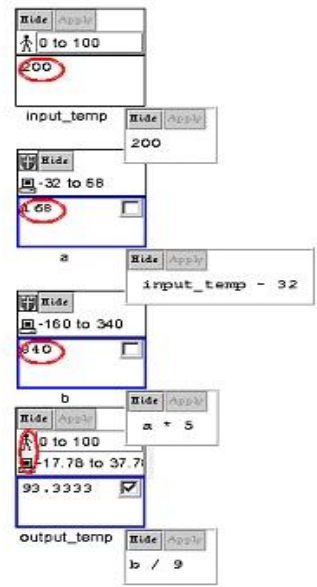

The approach provides three states for assertions to help end users detect faults in their spreadsheets. The first is when there is a conflict between a *USA* and an *SGA*, termed an *assertion conflict*, as in the oputput_temp cell in Figure 4, in which case the system will circle the conflict in red. The second is when the current value of a cell conflicts with the cell's assertions, termed a *value violation*.

Figure 4: An assertion conflict in Forms/3. [28]

For example, whenever the input_temp does not fall between 0 and 100, the system will circle it in red. Third, an *SGA* might look wrong to the end user. An empirical study [28] shows that an assertions-based approach significantly increases end users' effectiveness at testing and debugging spreadsheets' formulas. Assertions also help end users to detect a wide range of fault types, such as if an end user skips a step in a solution, non-reference faults, or structural errors [29].

Researchers have provided different approaches that help end users detect and validate formulas and numeric data. A study shows that almost 40% of spreadsheet cells contained either nonnumeric or non-date textual data [30]. Furthermore, a contextual inquiry revealed that information workers created nearly 70% of spreadsheets for reporting purposes [31]. Despite the importance of textual data, spreadsheets do not support validation strings. Though Microsoft Excel allows end users to associate data categories, such as phone numbers, with cells for formatting, it still does not validate the data values. As a result, an approach called Topes has been developed to allows end users to create and debug reusable, flexible data formats through a direct-manipulation user interface [32].

This technique relies on an abstraction called a tope that describes the formats and the relationships among those formats. An end user defines formats string data, which can be used to validate the data values in spreadsheets. For example, an end user might create a tope describes the format for phone numbers, and another may describe the format of URLs. Secondly, Topes uses these custom formats and highlights any cell that violates the format and generates two different functions, which are *isa* and *trf* from a tope description. The *isa* returns a number between 0 and 1 that indicates the degree to which a certain input string is an instance of the format. On the other hand, the *trf* function reformats the input string from one format to another target format only when *isa* function returns a value between 0 and 1. As a result, Topes allows for the definition of soft constraints about data that are true but not always true. For example, if an input data violates one of soft constraints, Topes displays a warning message that allows the end user to either accept or deny the input. Therefore, Topes helps end users to detect and fix different types of typo faults in string data.

To help end users in reducing the cognitive work of defining formats, Topes allows them to provide unlabeled textual examples and then infers data type formats and enables end users to review and customize the inferred formats [33].

### 3.2.3 Consistency checking

As discussed above, auditing approaches help end users identify outlier cells based on the hidden dependences between spreadsheets' cells whereas, testing approaches help end users make sure a spreadsheet is thoroughly tested and guide them in their search for errors. In this sub-section, I am going to review other approaches for automatically detecting errors.

Erwig and Burnett introduce a formal reasoning unit and identify rules that formalize the reasoning about the correctness formulas [34]. They use column headers that end users entered to label the data in their spreadsheets as unit declarations and use this information to carry out consistency checking of the spreadsheet formulas. The units are basically defined from the names of row and column headers in spreadsheets, and they can be formed into three different types. First are *dependent units*, which define all the hierarchical structure of units. For example, in the spreadsheet in Figure 5, the cell B3 is not just Apple, but Fruit [Apple]. Second, *and units* are defined by cell values that can be classified into different categories at the same time. In our example, C4 has the unit Fruit[Orange]&Month[June]. Third, *or units* are inferred for cells that contain operations combining values that have different units. For example, the unit of D3 gives an or unit of the units B3 and C3, which is Fruit[Apple]&Month[May]|Fruit[Orange]&Month[May], which it simplifies to a *well-formed unit*, which is Month[May] &Fruit[Apple|Orange]. Whenever simplification to a well-formed unit is not possible, a unit error is detected.

In addition, a visual system is implemented to support the formal reasoning and proposed mechanism that allows the end users to customize the inference mechanism described in [35].

Figure 5: Some headers for the Harvest spreadsheet. [34]

Yanif et. al. describe an approach that is based on the same principles as in [34], which they called "Unit Checking" [36]. However, there is a significant difference between two types of relationships between cells and their headers. The *is-a relationship* indicates for both instances and subcategories, and its structure is denoted with square brackets. The *has-a relationship* generally describes properties of items or sets, and its component is denoted by braces. The authors claim that their definition of relationships among the units allows their system to handle more errors in cell formulas than in [34]. For example, in Figure 6, the operation in cell B15, by following the rules in [34], would ordain the cells B3 and B9 have units **All Electronics [TVs[Gross]] & Year[2001]** and **All Electronics [VCRs[Gross]] & Year[2001]**, respectively. As a result, B19 has an error because its unit is not well formed because **TVs[Gross]** and **VCRs[Gross]** do not share a common prefix. On the other hand, Unit Checking describes cell B15 in terms of *is-a* and *has-a* relationships as **all Electronics [TVs]{Gross} &[Year[2001]] + all Electronics [VCRs]{Gross} &[Year[2001]]= all Electronics {Gross}.**

Figure 6: Electronics Sales and Profits. [36]

A principle problem with both previous approaches [35,36] is that they basically rely on the end user to annotate the value cells with the right units. Therefore, an automatic header inference approach was provided to make unit inference work in practice.

Since the unit inference depends on header information in a spreadsheet, an approach is provided called UCheck [37] that automatically infers the header and applies the reasoning and label normalization rules [38,34].

UCheck classifies spreadsheet cells into the following groups:
1- *Header*: the cells that contain strings that label the data.
2- *Footer*: usually cells at the end of rows or columns containing some sort of aggregation formula.
3- *Core*: the data cell.
4- *Filler*: empty or blank cells that separate tables in a spreadsheet.

Inferring and identifying the header is inherently a complicated task because end users use different styles of placing labels and headers in a spreadsheet. Therefore, UCheck implements different algorithms to detect a special arrangement of spreadsheet cells and classify them. Once header cells have been identified, UCheck then simplifies the unit information to a well-formed unit. If a formula does not have a well-formed unit, a unit error is reported. For example, in Figure 5, the headers of B2, B3, and B4 are Fruit, Month[May]&Fruit[Apple], and Month[June]&Fruit[Apple], respectively, and assuming that the formula in B5 is SUM(B2:B4), UCheck infers for B5 the header Fruit | Month[May]&Fruit[Apple] | Month[June]&Fruit[Apple], which is not well-formed since two units (B3 and B4) are compatible with each other but not with the first unit (B2). This error results from the incorrect range in the formula. As a result, this approach helps end users to detect errors in spreadsheet formulas, such as wrong references and illegal or omitted components in aggregation formulas.

Consistency checking approaches identify certain types of errors, such as reference, range, or omission errors, and report back errors by coloring cells that contain unit problems in their formulas, or cells that use a reference to another cell with a unit problem in their formulas.

Other researchers have integrated reasoning about the labels and interpreted them as units of measurements to provide information about the data being combined in formulas. Different approaches [39,41] have been developed to check the dimensions of formula cells for any inconsistencies. Dimensions are basically units of measurements expressed as types to the end users. For example, for a dimension such as length, there are several different units of measurements, such as cm, ft, m, or km [41].

Spreadsheet Language for Accentuating Type Errors (SLATE) is an approach that detects errors using dimension information provided by the end users [39]. SLATE requires the end user to annotate each cell with three attributes: a value, a unit, and a label. For example, a cell referring to 25 pounds of apples is annotated as (25, lbs., apples). Then, SLATE checks the formulas and infers dimensions and labels for these cells and reveals errors by displaying additional information in the cells: in addition to displaying a unit, it displays a label. For example, cell A2 is annotated (0.45, lb., apples), cell B2 is annotated (0.5, lb., oranges), cell A5 is annotated (312, lb., apples), and A6 is annotated (399, lb., oranges). The formula A2*A6 incorrectly computes the revenue from the sale of oranges by multiplying the weight of apples. The derived label for B6 reflects this error and has attributes from both apples and oranges.
$$(\$0.45, lb., apples) (399, lb., oranges) = \$179.55 (apples, oranges)$$

One major obstacle of using this approach is that the end user must type unit and labels as a text, which is a practice prone to errors. As a result, the end user might make a mistake while entering the unit information. To this end, another approach has been developed that infers the dimension information automatically rather than requiring the end user to annotate it [41]. Dimension analysis of spreadsheets is divided into the following phases.
1- Header inference, which identifies headers for each cell. The same technique as in [37,38] has been implemented for header inference.
2- Label analysis, which derives dimensions from each header according to the following process: (a) splitting headers into separate words, (b) extracting stems of the word, and (c) mapping word stems to dimensions, then combining them with each other.
3- Dimension inference, which inspects each cell's formulas and infers for it a dimension according to some specific rules [41]. Moreover, if these inferred dimensions are not the same with the dimensions that are extracted from the header, an error will be reported to the end user.

Reasoning approaches for inferring dimension information in spreadsheets help end users to identify errors in formulas, such as reference errors and wrong operands, because dimensions put constraints on how operations act on values.

# 4 Discussion and future works

In this paper, I presented a comprehensive review of different approaches that have been carried out to date to prevent, detect and remove errors from spreadsheets. Although they share a common goal, which is to help end users improve the quality of their spreadsheets without interfering with their goals, these approaches have not successfully either prevented or detected all types of errors that end users commonly make.

Most spreadsheet research focuses on ways to detect or decrease errors. Despite that, a variety of effective spreadsheet auditing approaches have been implemented, a major limitation of these approaches is that they do not measure the correctness of the spreadsheets and do not even provide any information on how to fix errors. Testing approaches, such as WYSIWYT and AutoTest, can help in discovering existing faults, but do nothing to remedy them. Another problem with testing approaches is that they might introduce more errors if the end users made incorrect decisions during testing. As a result, they might not solve the problem of complex spreadsheets. Although GoalDebug solves some of these problems by partly automating the testing/debugging process, it still relies on end users' decisions to give back good suggestions for fixing a faulty formula. Similarly, checking consistency using type inference approaches is limited by the kinds of errors they find. These types of approaches perform better on detecting errors if the end users spend time on the structuring of the spreadsheets.

Consequently, an empirical study has been conducted to combine the results from UCheck and WYSIWYT to improve fault localization for end users. The results of the study show the combinations of the two able to improve fault detection and they each can help the other overcome its limitations. In addition, the combination of feedback from UCheck and WYSIWYT is more effective than either approach alone [29].

There are other some areas where new techniques can be tried and incremental improvements can be made. One approach supports sharing and finding of topes for reuse, Sub-Section *3.2.2*. The approach uses machine-learning-based algorithms, such as collaborative filtering, to find appropriate topes in a repository and recommend them to end users for validating and reformatting the data while they are working with a spreadsheet. A laboratory experiment showed that the approach helps end users to validate and reformat spreadsheet cells twice as fast as manual editing with very low error rates [45]. Furthermore, other researchers have conducted three different lab experiments to understand the interactions between end users and an intelligent interface system, which sorts email messages into folders and explains its reasoning and behavior based on different algorithms [40]. The result shows that end users are willing to provide rich feedback to the system and it uses this feedback to improve its performance. In short comment, these results show evidence that there is a potential collaboration between end users and machine-learning systems.

One major limitation that spreadsheet systems have is the lack of abstractions of code reuse. A good way to highlight this limitation is to compare spreadsheet systems to programming languages. For example, in programming languages, such as Java, professional programmers abstract a block of code into a separate class, or method so that certain parts of their code can be repeatedly used. On the other hand, in spreadsheet programs when an end user wants to reuse a formula, he just copies it and pastes it to another region in the same or different spreadsheet. As a result, this leads to many types of referencing errors because many end users often miss updating the formulas with the new references.

As we have seen, a variety of approaches have been investigated to prevent errors from spreadsheets by providing end users with templates. These templates include all the formulas and the structure of the spreadsheet. ViTSL and ClassSheet approaches enable end users to infer templates from existing spreadsheets but not to recommend any templates to other end users.

Reuse and recommend templates or formulas among end users could have a number of benefits. First, reusing an existing template might be helpful for end users who are not familiar with a certain kind of organization pattern design for its spreadsheet. Second, it could be much more effective and efficient for end users to reuse an existing spreadsheet template than to create one from scratch. Third, recommending an existing template can reduce different types of errors, such as structural and reasoning errors. To this end, it is desirable to provide an automatic recommendation system through which end users can share, find, and reuse spreadsheet templates.

## 5 Conclusion

There is a large number of programs written by end users who do not have a solid background in computer science, and they use spreadsheets for their simplicity. However, with the increasing complexity of the spreadsheet, the development tasks become difficult and error-prone. Because of the prevalence of spreadsheet errors, there have been enormous efforts by researchers to produce approaches aimed at helping end users improve their spreadsheet quality and reliability. Spreadsheet research generally can be separated into three broad categories: prevention, detection, and removal of errors. Each of these reviewed approaches has its weaknesses, strengths, and types of errors it detects. Finally, the growing number of end users requires changing the direction of research to reduce spreadsheet errors by using and/or applying some machine-learning-based algorithms, such as collaborative filtering, for helping end users to find and reuse pre-existing templates.